\documentclass[a4paper,11pt,preprintnumbers]{article}

\usepackage{fullpage}
\usepackage[T1]{fontenc} 

\usepackage{amssymb,amsmath}

\begin{document} 
	
\begin{titlepage}
	\thispagestyle{empty}
	\begin{flushright}

		\hfill{DFPD-2016/TH/19}
	\end{flushright}

	\vspace{35pt}
	
	\begin{center}
	    { \LARGE{\bf Minimal Constrained Supergravity}}
		
		\vspace{50pt}

		{N.~Cribiori$^{1,2}$, G.~Dall'Agata$^{1,2}$, F.~Farakos$^{1,2}$} and M.~Porrati$^{3}$

		\vspace{25pt}

		{
			$^1$ {\it  Dipartimento di Fisica ``Galileo Galilei''\\
			Universit\`a di Padova, Via Marzolo 8, 35131 Padova, Italy}

		\vspace{15pt}

			$^2${\it   INFN, Sezione di Padova \\
			Via Marzolo 8, 35131 Padova, Italy}

		\vspace{15pt}

			$^3${\it   Center for Cosmology and Particle Physics, \\ 
			Department of Physics, New York University, \\ 
			4 Washington Place, New York, NY 10003, USA}

		}

		\vspace{40pt}

		{ABSTRACT}
	\end{center}
	
	\vspace{10pt}
	
	We describe minimal supergravity models where supersymmetry is non-linearly realized via constrained superfields. 
	We show that the resulting actions differ from the so called ``de Sitter'' supergravities because we consider constraints eliminating directly the auxiliary fields of the gravity multiplet.

\bigskip

\end{titlepage}

\baselineskip 6 mm

\section{Introduction} 
\label{sec:introduction}

If supersymmetry is realized in nature, it is likely that its breaking scale is much higher than the one currently probed experimentally.
It may also be as high as the string scale.
In a scenario of this type, it is then difficult to find a good use to the phenomenological models where supersymmetry has a linear realization on the fields describing elementary particles and their interactions.
However, also at energy scales much lower than the supersymmetry breaking scale, supersymmetry could pose visible constraints to interactions, especially if the mass spectrum is split so that are some states that are much lighter than other ones.
For this reason, it is interesting to study non-linear realizations of supersymmetry and understand how to construct a general formalism that can be efficiently used to implement them in phenomenological models.

While non-linear realizations have been already studied shortly after the introduction of supersymmetry \cite{Volkov:1973ix,others}, it is only in the last couple of years that they received a broader attention.
This is especially for their possible application to cosmological models (early references on this subject are \cite{Antoniadis:2014oya,Ferrara:2014kva,Kallosh:2014via,Dall'Agata:2014oka}, while \cite{Ferrara:2015cwa,Ferrara:2016ajl} provide recent reviews).
The recent resurgence of interest in the subject requires a study of non-linear realizations in the context of supergravity theories, that revisits and extends the results obtained in global supersymmetry.

An important step has been the construction of non-linear models of pure supergravity \cite{Antoniadis:2014oya,Dudas:2015eha,Bergshoeff:2015tra,Hasegawa:2015bza}.
These are models where the physical spectrum contains only the graviton, the gravitino and the goldstino.
While these models have been dubbed ``de Sitter'' supergravities, the cosmological constant depends on two parameters, related to the susy-breaking scale $f$ and to the gravitino mass $m_{3/2}$, so that depending on their value it can be positive, negative or vanishing.
These models effectively couple the supergravity multiplet, where supersymmetry is linearly realized, to the goldstino, which is the goldstone field of supersymmetry breaking and provides a non-linear realization of the supersymmetry algebra.
After integration of the auxiliary fields and in the unitary gauge, where the goldstino is set to zero, the component action for these models is described by 
\begin{equation}\label{eq:desittersugra}
		e^{-1} {\cal L} = -\frac12 \, R
		+ \frac{1}{2} \epsilon^{klmn} (\overline \psi_k \overline \sigma_l {\cal D}_m \psi_n - \psi_k \sigma_l {\cal D}_m \overline\psi_n) 
		- (m_{3/2} \, \overline \psi_a \overline \sigma^{ab} \overline \psi_b 
		+ \overline m_{3/2} \, \psi_a \sigma^{ab} \psi_b ) - \Lambda,
\end{equation}
where $\Lambda= |f|^2 - 3 |m_{3/2}|^2$ is the cosmological constant, which is determined by the only two parameters in the theory: the susy-breaking scale $f$ and the gravitino mass $m_{3/2}$.
Throughout this paper we use reduced Planck mass units that set $8\pi G=1$ unless explicitly stated otherwise.

In this note we will construct minimal supergravity models, whose physical spectrum is also given by the graviton, the gravitino and the goldstino, but where supersymmetry is non-linearly realized already on the gravity multiplet itself.
We will see that this produces interactions and Lagrangians that differ from those in \cite{Antoniadis:2014oya,Dudas:2015eha,Bergshoeff:2015tra,Hasegawa:2015bza} and which depend on three independent physical inputs: the susy-breaking scale, the gravitino mass and the cosmological constant.
This problem has been already tackled from a different perspective in \cite{Delacretaz:2016nhw}, with the purpose of constructing an effective field theory for supergravity models of inflation.
Ref.~\cite{Delacretaz:2016nhw} uses a supersymmetric generalization of the CCWZ construction \cite{Callan:1969sn}. 
We will instead perform our analysis in the language of superfields allowing for more general and different constraints.
In fact, it is known that non-linear realizations of supersymmetry can be obtained by imposing supersymmetric constraints to superfields, so that some of their component fields get removed from the spectrum.
While for a long time these constraints were imposed on a case by case basis, we now have a general procedure \cite{Dall'Agata:2016yof}, which can be used to consistently remove any unconstrained field from a multiplet, both in global and local supersymmetric theories.
In the following we are going to apply this procedure to remove the auxiliary fields of the minimal supergravity multiplet, detailing the features of the resulting models.

We will show that the most general models we can construct depend on 3 parameters, related to the scale of supersymmetry breaking $\Lambda_S$, to the gravitino mass $m_{3/2}$ and to the cosmological constant $\Lambda$.
Already in the unitary gauge we can see that these models differ from those given in \cite{Antoniadis:2014oya,Dudas:2015eha,Bergshoeff:2015tra,Hasegawa:2015bza}, which are described by (\ref{eq:desittersugra}).
In fact, after integrating any eventual auxiliary fields and in the unitary gauge, the models we obtain have a cosmological constant given by
\begin{equation}
	\Lambda = \frac13 |c|^2+ |f|^2 + m_{3/2} \overline c +\overline m_{3/2} c = \Lambda_S - 3 |m_{3/2}|^2\,,
\end{equation}
where $f$ is the $F$-term of the goldstino multiplet and $c$ is a new parameter which can be introduced when constraining the supergravity auxiliary scalar field.


\section{Minimal constrained supergravity} 
\label{sec:minimal_constrained_supergravity}

In the language of superfields, the minimal supergravity multiplet is described by means of two different superfields ${B}_{\alpha \dot \alpha}$ and ${\cal R}$, related by algebraic constraints.
Using the conventions of \cite{Wess:1992cp}, to build the Lagrangian we also need the chiral density ${\cal E}$. Its expansion in components is
\begin{equation}
	\label{calE}
	2 {\cal E} = e \left\{1 + i\,\Theta \sigma^a \overline\psi_a - \Theta^2 (m^* + \overline\psi_a \overline{\sigma}^{ab} \overline\psi_b)\right\},
\end{equation}
where $e$ is the determinant of the vielbein, $\psi_a$ is the gravitino and $m$ is the complex auxiliary scalar field.
The curvature superfield ${\cal R}$ is also a chiral superfield 
\begin{equation}
	\label{calR}
	\begin{array}{rcl}
	 	{\cal R} &=&  \displaystyle - \frac16 \left\{m + \Theta(2\sigma^{ab} \psi_{ab} - i \sigma^a \overline\psi_a \,m + i \psi_a b^a)\right.  \\[3mm]
		&+& \displaystyle \Theta^2\left(-\frac12 R + i \overline \psi^a\overline{\sigma}^b \psi_{ab} + \frac23 |m|^2 + \frac13 b^2 - i\, {\cal D}_a b^a \right. \\[3mm]
		&+& \displaystyle \left. \left. \frac12 \overline{\psi}\overline{\psi}\,m-\frac12 \psi_a \sigma^a \overline\psi_c \,b^c + \frac18\, \epsilon^{abcd}(\overline\psi_a \overline \sigma_b \psi_{cd} + \psi_a \sigma_b \overline \psi_{cd})\right)\right\}.
	\end{array}
\end{equation}
It contains the gravitino field-strength, the Ricci scalar curvature $R$ and the auxiliary vector field $b_a$.
The real superfield $B_{\alpha \dot \alpha}$ has the auxiliary vector field $b_a$ as lowest component and it is related to ${\cal R}$ by means of the superspace Bianchi identity
\begin{equation}
	\label{Bianchi}
	{\cal D}^{\alpha} B_{\alpha \dot \alpha} = \overline {\cal D}_{\dot \alpha} \overline{\cal R}.
\end{equation}
Supersymmetry breaking requires also the existence of a goldstino field, which can be described by means of a chiral superfield $X$, constrained by the nilpotency condition $X^2 = 0$ \cite{others,Casalbuoni:1988xh}.
The latter solves the scalar field in the lowest component in terms of the goldstino ${\cal G}$
\begin{equation}
	\label{Xexp}
	X = \frac{{\cal G}^2}{2 F} + \sqrt2\, \Theta {\cal G} + \Theta^2 F,
\end{equation}
provided ${\cal D}^2 X = -4 F$ is non-zero on the vacuum.
The models discussed in \cite{Antoniadis:2014oya,Dudas:2015eha,Bergshoeff:2015tra,Hasegawa:2015bza} correspond to the coupling of this superfield to the unconstrained supergravity multiplet.

Here we decided to follow a different path and impose supersymmetric constraints on the supergravity superfields in order to remove the auxiliary fields $m$ and $b_a$.

The scalar auxiliary field $m$ can be removed by imposing the constraint
\begin{equation}
	\label{XR}
	X {\cal R} = 0,
\end{equation}
which fixes the lowest component of ${\cal R}$ in terms of a function of the goldstino, the gravitino, the Riemann curvature and the other auxiliary field $b_a$.
This constraint on chiral superfields has been first introduced in global supersymmetry in \cite{Brignole:1997pe} and then applied to matter superfields in supergravity in \cite{Dall'Agata:2015zla}.
Here we apply it directly to the supergravity curvature superfield and therefore we use it to constrain an auxiliary field.
We will see that such constraint, as noted in \cite{Dall'Agata:2015lek} for other constraints on auxiliary fields, implies that the final form of the potential is going to be different from the one of standard supergravity models.
For this reason it is actually interesting to impose a slightly different constraint, namely
\begin{equation}
	\label{XRc}
	X \left({\cal R} + \frac{c}{6}\right) = 0.
\end{equation}
For a generic chiral superfield this constraint simply adds a constant vacuum expectation value to the scalar field in the ${\cal R}$ multiplet, but for the supergravity curvature superfield this has a non-trivial implication on the effective Lagrangian.
As usual, we can consistently solve (\ref{XRc}) by inspecting its top component.
Given the peculiar structure of the ${\cal R}$ superfield, which contains the auxiliary field $m$ in various places, its solution can be found only after iteratively applying the constraint on $m$.
The final result is 
\begin{equation}
	\label{msol}
	\begin{array}{rcl}
		m &=& \displaystyle N\left(1-\frac{i}{\sqrt2 F} {\cal G}\sigma^a \overline \psi_a\right) + c\left(1-\frac{i}{\sqrt2 F} {\cal G} \sigma^a \overline \psi_a-\frac{1}{2 F^2}({\cal G} \sigma^b \overline\psi_b)^2-\frac{1}{4F^2} \overline\psi\,\overline\psi\,{\cal G}{\cal G}\right)\\[5mm]
		&-&\displaystyle \frac{c}{3 F^2} \,{\cal G}{\cal G}\left[\overline N\left(1-\frac{i}{\sqrt2 \overline{F}}\overline{\cal G}\,\overline\sigma^a\psi_a\right)+\overline{c}\left(1-\frac{i}{\sqrt2 \overline{F}} \overline{\cal G}\,\overline\sigma^a\psi_a-\frac{1}{2 \overline F^2}(\overline{\cal G}\,\overline \sigma^b\psi_b)^2-\frac{1}{4 \overline F^2} \overline{\cal G}\overline{\cal G} \,\psi \psi\right)\right.\\[5mm]
		&-& \left. \displaystyle\frac{|c|^2}{3 \overline F^2}\, \overline{\cal G}\overline{\cal G}\right],
	\end{array}	
\end{equation}
where
\begin{equation}
\label{NIN}
	\begin{array}{rcl}
		N &=& \displaystyle\frac{\sqrt2}{F} {\cal G} \sigma^{ab}\psi_{ab} + \frac{i}{\sqrt2 F} b^a \,{\cal G}\psi_a 	+\frac{1}{4 F^2} {\cal G}{\cal G} {\cal R} 	+ \frac{i}{2 F^2} {\cal G}{\cal G} {\cal D}_a b^a \\[5mm]
		& +&\displaystyle \frac{1}{2 F^2} {\cal G}{\cal G}\left[\frac12 \psi_m \sigma^m \overline\psi_n b^n-\frac13 b^2-i\, \overline\psi^m \overline\sigma^n \psi_{mn}-\frac18\epsilon^{klmn}\left(\overline\psi_k\overline\sigma_l\psi_{mn}+\psi_k\sigma_l\overline\psi_{mn}\right)\right]. 
	\end{array}
\end{equation}

The minimal supergravity model with non-linear supersymmetry, subject to the constraint (\ref{XRc}), is defined by the K\"ahler potential
\begin{equation}
	\label{K\"ahler}
	K = X \overline{X}
\end{equation}
and the superpotential
\begin{equation}
	\label{W}
	W = m_{3/2} + f \, X.
\end{equation}
Using the nilpotency of $X$ and the constraint on ${\cal R}$, we see that the generic superspace Lagrangian \cite{Wess:1992cp}
\begin{equation}
	\label{supL}
	{\cal L} = \frac{3}{8} \int d^2 \Theta \,  2 {\cal E} \,  (\overline {\cal D}^2 - 8 {\cal R} )  
	e^{- \frac{1}{3} K(X,\overline X)}+ \int d^2 \Theta \,  2 {\cal E} \, W + h.c.,
\end{equation}
reduces to the sum of three terms: the Einstein--Hilbert action
\begin{equation}
	\label{L01}
	{\cal L}_1 = - 6 \int d^2 \Theta \, {\cal E} \,  {\cal R} +h.c.,
\end{equation}
the K\"ahler potential\footnote{Recall that in supergravity the  operator $-1/4 (\overline{\cal D}^2 -8{\cal R})$ transforms an antichiral field into a chiral field.} for the nilpotent field $X$ 
\begin{equation}
	\label{L02}
	{\cal L}_2 = \int d^2 \Theta \, {\cal E} \,  X \left[-\frac14 \left(\overline{\cal D}^2 -8{\cal R}\right)\right]\overline{X} + h.c.
\end{equation}
and the superpotential
\begin{equation}
	\label{L03}
	{\cal L}_3 = \int d^2 \Theta \,  2 {\cal E} \, W + h.c.\,.
\end{equation}
The associated component Lagrangian in the unitary gauge, where ${\cal G} = 0$, is
\begin{equation}\label{unitaction}
	\begin{array}{rcl}
		e^{-1} {\cal L} &=& \displaystyle  -\frac12 \, R
		+ \frac{1}{2} \epsilon^{klmn} (\overline \psi_k \overline \sigma_l {\cal D}_m \psi_n - \psi_k \sigma_l {\cal D}_m \overline\psi_n) + \frac13 b_a b^a
		 \\[5mm]
		&-&  \displaystyle (m_{3/2} \, \overline \psi_a \overline \sigma^{ab} \overline \psi_b 
		+ \overline m_{3/2} \, \psi_a \sigma^{ab} \psi_b ) - \Lambda,
	\end{array}
\end{equation}
where the cosmological constant $\Lambda$ is 
\begin{equation}
	\label{potential}
	\Lambda = \frac13 |c|^2+ |f|^2 + m_{3/2} \overline c +\overline m_{3/2} c = \Lambda_S - 3 |m_{3/2}|^2\,,
\end{equation}
where we introduced the effective supersymmetry breaking scale 
\begin{equation}
	\label{LambdaS}
	\Lambda_S = |f|^2 + \left|\frac{c}{\sqrt3} + \sqrt3\, m_{3/2}\right|^2.
\end{equation}
We see that we have three independent parameters in the Lagrangian, corresponding to the gravitino mass $m_{3/2}$, the scale of supersymmetry breaking $\Lambda_S$ and the cosmological constant $\Lambda$.
Action~(\ref{unitaction}) depends only on two independent functions of these three parameters, but an action describing supergravity coupled to matter would depend generically on all three parameters. 
As announced, the cosmological constant does not take the standard form of an ordinary supergravity potential, since the latter is given by the difference of the $F$-terms squared minus (3 times) the squared gravitino mass.
The constraint on the supergravity scalar auxiliary field implies a different form, whose numerical value can be positive, negative or zero, depending on the choice of the parameters $f$, $m_{3/2}$ and $c$.
When $c=0$ we obtain a pure de Sitter supergravity, with a cosmological constant $\Lambda = |f|^2$.
When $c = -3 {W}_0$ on the other hand, we obtain a cosmological constant in the standard form $\Lambda = |f|^2 - 3 |m_{3/2}|^2$ and the supersymmetry breaking scale directly related to $f$, i.e. $\Lambda_S = |f|^2$. 
This is for instance what one expects from models where supersymmetry is realized linearly on the gravity multiplet and the auxiliary field $m$ gets replaced by its equation of motion $m = -3\, m_{3/2} + \ldots$\,.

In the construction above, the vector auxiliary field has been left untouched, but we could also impose a further constraint to eliminate it from the spectrum.
Since the superfield that contains $b_a$ as its lowest component is a real vector field, we need to use the prescription of \cite{Dall'Agata:2016yof} and impose the constraint
\begin{equation}
	\label{Bconst}
	X \overline{X} B_{\alpha \dot \alpha} = 0.
\end{equation}
This can be consistently done because $B_{\alpha \dot \alpha}$ contains $b_a$ nakedly and not through its field-strength.
As demonstrated in \cite{Dall'Agata:2016yof}, this constraint has a unique solution for the lowest component, which can be obtained as the $\theta = \bar \theta = 0$ projection of
\begin{equation}
B_a = -2 \frac{\overline{\cal D}_{\dot\alpha}\overline X}{\overline{\cal D}^2 \overline X}\overline{\cal D}^{\dot \alpha}B_a - \overline X \frac{\overline{\cal D}^2 B_a}{\overline{\cal D}^2\overline X}-2\frac{{\cal D^\alpha}X}{{\cal D}^2 X\,\overline{\cal D}^2\overline X} {\cal D}_\alpha \overline{\cal D}^2(\overline X B_a) -\frac{X}{{\cal D}^2 X \overline{\cal D}^2 \overline X}{\cal D}^2 \overline{\cal D}^2 (\overline X B_a).
\end{equation}
This produces an expression for $b_a$ that is a function of the goldstino, the gravitino and the graviton, in a way that it vanishes in the unitary gauge.
Clearly, the addition of this constraint further complicates the expression of the other auxiliary field and modifies the couplings of the goldstino in the final Lagrangian.


\section{Non-linear realizations and constrained superfields} 
\label{sec:non_linear_realizations_and_constrained_superfields}

As we mentioned in the introduction, the authors of \cite{Delacretaz:2016nhw} also discuss non-linear realizations of minimal supergravity models, though in a different formalism and with a different purpose.
We will now comment on the relation between the two approaches and the differences in the results.

The first thing to note is that \cite{Delacretaz:2016nhw} uses a parametrization for the goldstino field that is different from the one obtained by using the constrained superfield $X$.
As it is known, there are actually various different non-linear realizations of the same superalgebra.
The one of \cite{Delacretaz:2016nhw} uses the original Volkov--Akulov formulation \cite{Volkov:1973ix}, where the goldstino $\lambda$ transforms under supersymmetry as
\begin{equation}
\delta \lambda_\alpha = - f \, \epsilon_\alpha + \frac{i}{f} \left( \lambda \sigma^m \overline \epsilon - \epsilon \sigma^m \overline \lambda  \right) {\cal D}_m \lambda_\alpha + \cdots .
\end{equation} 
The scale of susy-breaking $f$ was set to one in \cite{Delacretaz:2016nhw}.
A different formulation is that of Samuel and Wess \cite{Samuel:1982uh}, where the goldstino $\gamma$ is related to the one of Volkov--Akulov by a non-trivial field redefinition (which can be found in \cite{Liu:2010sk,Kuzenko:2010ef} for global supersymmetric theories) and transforms as
\begin{equation} 
\delta \gamma_\alpha = - f\, \epsilon_\alpha + 2 \frac{i}{f}\,  \gamma \sigma^m \overline \epsilon \, {\cal D}_m \gamma_\alpha + \cdots. 
\end{equation} 
The one that comes from the constrained superfield representation is then simply a linear field redefinition, where
\begin{equation}
\label{gamma}
G_{\alpha} = \sqrt 2 \, \frac{F}{f}\, \gamma_{\alpha}.
\end{equation}
Once we have this realization, we can see that we can translate the constraints imposed in \cite{Delacretaz:2016nhw} in terms of constrained superfields using the general recipe given in \cite{Dall'Agata:2016yof}.
Following \cite{Delacretaz:2016nhw}, starting from a given component $\phi$ of a supermultiplet, one can impose a supersymmetric constraint by imposing
\begin{equation}
\hat \phi =\left( D_G \Big{[} \text{e}^{\epsilon^\alpha \hat Q_\alpha} \Big{]} \times \phi \right)_{\epsilon=\lambda} = \phi + \delta_\epsilon \phi |_{\epsilon=\lambda} 
+ \frac12 \delta_\epsilon \delta_\epsilon \phi |_{\epsilon=\lambda} + \cdots =0,
\end{equation}
where $D$ is some representation of the group $G$ (which in our case is the supergroup), while $\hat{Q}$ is the infinitesimal generator of the symmetry transformation with parameter $\epsilon(x) = \lambda(x)$. 
This equation results in a constraint that removes the field $\phi$ from the spectrum.
In the language of constrained superfields one can impose the same condition for a generic multiplet $\Phi$, whose lowest component is $\phi$ by setting \cite{Dall'Agata:2016yof}
\begin{equation}
\label{TXC}
X \overline X \Phi = 0.
\end{equation}
This eliminates the lowest component field of the superfield $\Phi$, namely $\phi$, while also inducing a non-linear realization of supersymmetry for the remaining component fields. 
It is straightforward to see that the constraint (\ref{TXC}) implies 
\begin{equation}
\label{txc}
({\cal G}{\cal G})  (\overline{\cal G}\overline{\cal G}) \phi = 0,
\end{equation} 
which produces the full constraint \eqref{TXC} once we require that this condition be invariant under supersymmetry.  
Indeed if the lowest component of a superfield vanishes, the whole superfield vanishes. 
The two formulations give the same result, because the solution of the constraint $\hat{\phi} = 0$ implies 
\begin{equation}
	(\lambda \lambda)(\overline \lambda \overline \lambda) \phi = 0,
\end{equation}
which, using the field redefinition from $\lambda$ to ${\cal G}$, is equivalent to (\ref{txc}).
We therefore conclude that the results of \cite{Delacretaz:2016nhw} correspond to imposing the constraints (\ref{XR}) and (\ref{Bconst}) in the constrained superfields language.
One can actually be convinced that this is the case by direct computation of the solution of the constraints of \cite{Delacretaz:2016nhw}.
For instance, the constraint $\hat{m} = 0$ gives
\begin{equation}
\begin{aligned}
\label{Mgeom}
m &= -\delta_\lambda m +\cdots
&=2\lambda \sigma^{ab}\psi_{ab}+ib^a\lambda \psi_a-2i(\lambda\sigma^a\overline\psi_a)(\lambda \sigma^{ab}\psi_{ab})+\cdots
\end{aligned}
\end{equation}
which agrees with the first terms in \eqref{msol} after using the relation between ${\cal G}$ and $\lambda$.

This comparison also shows that imposing the constraint (\ref{XRc}) we produce a Lagrangian that differs from the one in \cite{Delacretaz:2016nhw}. 
In fact, if we do not introduce the parameter $c$, we see from (\ref{potential}) that the cosmological constant can only be positive and proportional to the susy-breaking parameter $f$.
Actually, even the coupling of additional matter multiplets would not change this fact, leading to a true ``de Sitter'' supergravity.
The only way to change this fact would be by changing the sign and the overall coefficient to $X \overline X$ in (\ref{L02}), which is the term supersymmetrizing the cosmological constant term.
If $X$ were a standard chiral superfield this fact would produce a ghost from the scalar component $x$, but here we have a constrained superfield and $x$ is replaced by a bilinear of the goldstino.
Still, we expect that introducing this change in sign in front of this term would tell us that any ultraviolet completion of such a  model would be pathological.


\section{Discussion} 
\label{sec:discussion}

In this letter we considered minimal supergravity models where supersymmetry is non-linearly realized by constraining the auxiliary fields of the supergravity multiplet.
The resulting theories depend on three parameters related to the cosmological constant, the mass of the gravitino and the supersymmetry breaking scale.
All these three parameters appear in independent combinations in a generic matter-coupled supergravity but the pure broken supergravity constructed in this letter depends of course only on two independent parameters: the gravitino mass $m_{3/2}$ and the cosmological constant $\Lambda$.
This is equivalent to say that the actions constructed in this note  
differ off-shell from those appearing in the literature, either because the latter had supersymmetry realized linearly on the gravity fields, or because the constraints imposed on the auxiliary fields were chosen differently.

An obvious interesting development is the coupling of these models to matter fields, either free or constrained.
Already in the case where supersymmetry is linearly realized on the matter fields we see nontrivial effects of the new constraints imposed on the gravity multiplet.
Since we impose a constraint on the auxiliary field $m$, the scalar potential cannot be put in the standard supergravity form, where the negative contribution, proportional to the superpotential, comes precisely by the integration of $m$.
Also, if one imposes the constraint (\ref{Bconst}) on the vector auxiliary field, we expect that the conditions on the scalar manifolds get relaxed.
In particular, we expect that the scalar $\sigma$-model will be described by an arbitrary K\"ahler manifold rather than by a more restrictive Hodge--K\"ahler manifold.
In fact, while the equations of motion of the $b_a$ auxiliary field usually solve it in terms of the composite K\"ahler connection, the constraint (\ref{Bconst}) sets it to zero in the unitary gauge.
This should also imply that the fermions are trivial sections of the K\"ahler bundle.
If, in addition, we couple constrained matter fields, then even the K\"ahler structure of the scalar $\sigma$-model is lost.

Another observation worth making is that the constraints we proposed here can be used to give a natural embedding of the $R^2$ ``Starobinsky'' model of inflation in supergravity.
This can be obtained as a linear combination of the terms ${\cal L}_2$, ${\cal L}_3$, that are defined in eqs.~(\ref{L02},\ref{L03}) and include the goldstino action and the 
supersymmetry-breaking superpotential (where we set $m_{3/2} =0$ for simplicity), 
plus a term quadratic in the curvature
\begin{equation}
{\cal L}_{4} = 54\, \alpha \int d^2 \Theta (2{\cal E}) 
{\cal R} \left[-\frac14\,\left(\overline{\cal D} - 8 {\cal R}\right)\right] \overline {\cal R}  + c.c.\qquad \alpha=\mbox{constant}.
\end{equation}
 
Imposing the constraint $X^2 = 0$ and the curvature constraints (\ref{XRc}), (\ref{Bconst}), we break both conformal invariance and supersymmetry. We also get rid of the auxiliary fields in the gravity action thus producing a bosonic sector which is very simple:
\begin{equation}
	e^{-1} {\cal L} = - \frac12  |c|^2 \alpha R + \frac{3}{4} \alpha R^2 +
	\frac{\alpha}{3} |c|^4 -|f|^2.
\end{equation}
If we normalize the Einstein--Hilbert term as usual and require the vanishing of the cosmological constant, the parameters are constrained so that for high-scale susy-breaking $\sqrt{|f|} \sim 10^{-3} M_P$ we naturally produce the large coefficient $\alpha=O(10^5)$ needed to make the Starobinsky model consistent with CMB data.



\bigskip
\section*{Acknowledgments}

\noindent We would like to thank S.~Ferrara, A.~Kehagias, S.~Rigolin and A.~Wulzer for interesting discussions.
This work is supported in part by the Padova University Project CPDA119349, by the MIUR grant RBFR10QS5J \emph{``String Theory and Fundamental Interactions''}. MP is supported 
in part by NSF grants PHY-1316452 and PHY-1620039.
We acknowledge hospitality of the GGI institute in Firenze, where this work was performed during the workshop `Supergravity: what next?'. 


\end{document}